# MODELLING THE GALACTIC BAR USING RED CLUMP STARS

K. Z. Stanek[1]

*Princeton University Observatory, Princeton, NJ 08544–1001*

M. Mateo

*Department of Astronomy, University of Michigan,*
*821 Dennison Bldg., Ann Arbor, MI 48109–1090*

A. Udalski, M. Szymański, J. Kałużny, M. Kubiak

*Warsaw University Observatory,*
*Al. Ujazdowskie 4, 00–478 Warszawa, Poland*

W. Krzemiński

*Carnegie Observatories, Las Campanas Observatory,*
*Casilla 601, La Serena, Chile*

**Abstract.**
The color-magnitude diagrams of $\sim 1\times 10^6$ stars obtained for 19 fields towards the Galactic bulge with the OGLE project reveal a well-defined population of bulge red clump stars. We found that the distributions of the extinction-adjusted apparent magnitudes of red clump stars in fields lying at $l = \pm 5°$ in galactic longitude differ by $\sim 0.4$ *mag*. A plausible explanation of this observed difference in the luminosity distribution is that the Galactic bulge is a triaxial structure, or a bar, which is inclined to the line of sight by no more than $45°$. The part of the bar at the positive galactic longitude is closer to us. Work is now under way to model the Galactic bar by fitting the observed luminosity functions in the red clump region for various fields. Preliminary results indicate that the angle of the inclination of the bar to the line of sight can be as small as $\sim 20°$. Gravitational microlensing can provide us with additional constrains on the structure of the Galactic bar.

## 1. INTRODUCTION

The Optical Gravitational Lensing Experiment (OGLE, Udalski et al. 1994; Paczyński et al. 1995 and references therein) is an extensive photometric search for the rare cases of gravitational microlensing of the Galactic bulge stars by foreground objects. It provides a huge database (Szymański & Udalski 1993), from which color-magnitude diagrams have been compiled (Udalski et al. 1993, 1995). Here we discuss the use a of a well-defined population of bulge red clump star to investigate the presence of the bar in our Galaxy. The results of our earlier studies are described by Stanek et al. (1994, 1995a).

---

[1] On leave from N. Copernicus Astronomical Center, Bartycka 18, Warszawa 00–716, Poland

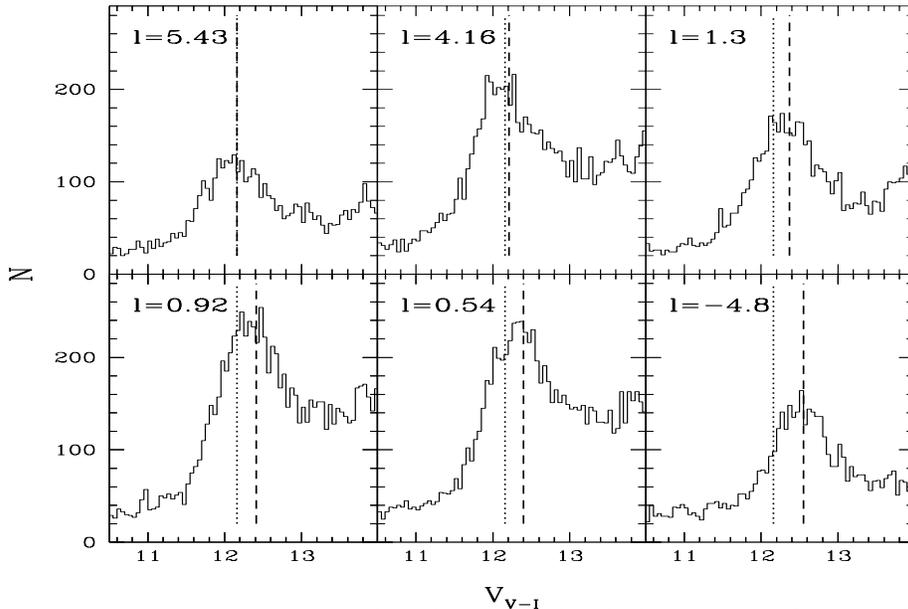

Figure 1. The $V_{V-I}$ distributions of red clump stars for six of our fields. With the vertical dashed lines we mark the position of the mode of the distributions and with the dotted line we show the mode of the first distribution. There is a clear anti-correlation between the $V_{V-I}$ value of the mode and the Galactic longitude $l$ for a given field. This indicates that the bulge is not axially symmetric.

There is now a number of photometric and dynamical indications that the Galaxy is barred (for a review see Kuijken 1995 – these proceedings). The bar is clearly present in the COBE DIRBE data, which was used by Dwek et al. (1995) to constrain a number of analytical bar models existing in the literature.

## 2. THE DATA

Udalski et al. (1993, 1995) present color-magnitude diagrams (CMDs) of 19 fields in the direction of the Galactic bulge, which cover nearly 1.5 square degrees and contain about $8 \times 10^5$ stars. Here we use a well-defined population of bulge red clump stars to investigate the presence of the bar in our Galaxy.

To analyze the distribution of bulge red clump stars in a quantitative manner, we define the extinction-insensitive $V_{V-I}$ parameter

$$V_{V-I} \equiv V - 2.6 \, (V - I), \qquad (1)$$

where we use reddening law $E_{V-I} = A_V/2.6$. The parameter $V_{V-I}$ has been defined so that if $A_V/E_{V-I}$ is independent of location then for any particular star its value is not affected by the unknown extinction (Stanek et al. 1994; Woźniak & Stanek 1995). For all 19 fields we select the region of the CMD clearly dominated by the bulge red clump stars. For every field all stars in that region were counted in bins of $\Delta V_{V-I} = 0.05$. The result for some of the fields appears in Fig.1, where we show the number of stars as a function of $V_{V-I}$. The fields were ordered from top-left to bottom-right by decreasing galactic longitude $l$. The distributions shown in Fig.1 are similar in shape, with red clump stars forming a pronounced peak. There is however a clear shift between the distributions, in the sense that stars from fields with bigger value of $l$ have on average smaller values of $V_{V-I}$ parameter. To quantify this shift, for every field we found the mode of the $V_{V-I}$ distribution (Fig.1). There is a clear anti-correlation between the $V_{V-I}$ value of the mode and the Galactic longitude $l$ for a given field that corresponds to decrease of $V_{V-I}$ value of $\sim 0.04\ mag/°$.

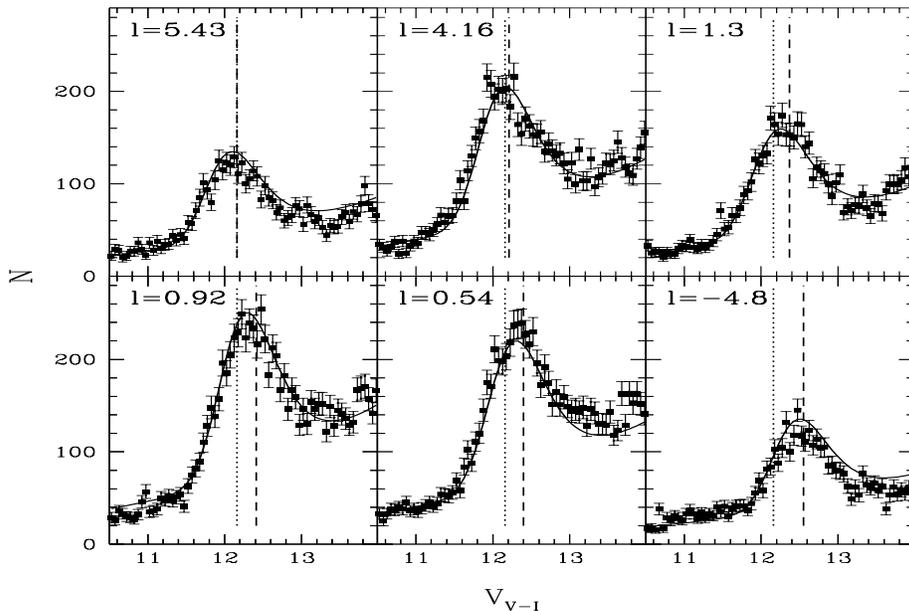

Figure 2. Model fit (thick continuous line) to the $V_{V-I}$ distributions of the same six fields as in Fig.1.

## 3. DISCUSSION

For various fields we have shown that the distributions of bulge red clump stars as a function of extinction-adjusted apparent magnitude are similar in shape but are systematically shifted. This is likely due to the difference in the distance to the bulge red clump stars in various fields and indicates that the bulge is not axially symmetric.

We use the observed luminosity function of red clump stars to put constraints on various models of the Galactic bar. This has the advantage over studies using surface brightness measurements (Dwek et al. 1995) that the red clump stars provide information about the depth of the bar along the line of sight. Such information was recently used by Stanek (1995) in his study of the magnitude offset between gravitationally lensed stars and observed stars. The preliminary results from the modelling of the Galactic bar by fitting the observed red clump luminosity function (Stanek et al. 1995b, Fig.2) indicate that the inclination of the Galactic bar to the line of sight is about $\sim 20°$. However, the current data is not sufficient to discriminate between various models of the Galactic bar 3-D mass distribution, so during the current OGLE observing season we are collecting data for additional fields across the bulge.

The presence of the bar in the Galaxy seems to be firmly established by various authors and methods, but there are still considerable differences as to details of the bar structure or angle of inclination to the line of sight. We have shown that the red clump stars can be very useful for investigating the Galactic bar, being both numerous and relatively bright. We are now extending the work presented here, by incorporating more information provided by the red clump region of CMDs, to test a variety of Galactic bar models (Stanek et al. 1995b).

**Acknowledgments.** We would like to thank B. Paczyński, the PI of the OGLE project, for encouragement, many stimulating discussions and comments. We also want to thank A. Ulmer for careful reading of this contribution. This work was supported with the NSF grants AST 9216494 and AST 9216830 and Polish KBN grants No 2-1173-9101 and BST438A/93. KZS thanks also for the NAS Grant-in-Aid of Research through Sigma Xi, The Scientific Research Society.

**References**

Dwek, E., et al. 1995, ApJ, 445, 716

Kuijken, K., 1995, these Proceedings

Paczyński, B., Stanek, K. Z., Udalski, A., Szymański, M., Kałużny, J., Kubiak, M., Mateo, M., Krzemiński, W. & Preston, G. W., 1995, in: IAU Symposium 169, "Unsolved Problems of the Milky Way", ed. L. Blitz, in press

Stanek, K. Z., 1995, ApJ, 441, L29

Stanek, K. Z., Mateo, M., Udalski, A., Szymański, M., Kałużny, J., & Kubiak, M., 1994, ApJ, 429, L73

Stanek, K. Z., Mateo, M., Udalski, A., Szymański, M., Kałużny, J., Kubiak, M., & Krzemiński, W., 1995a, in: IAU Symposium 169, "Unsolved Problems of the Milky Way", ed. L. Blitz, in press

Stanek, K. Z., et al. 1995b, in preparation

Szymański, M., & Udalski, A. 1993, Acta Astron., 43, 91

Udalski, A., Szymański, M., Kałużny, J., Kubiak, M., & Mateo, M. 1993, Acta Astron., 43, 69

Udalski, A., Szymański, M., Stanek, K. Z., Kałużny, J., Kubiak, M., Mateo, M., Krzemiński, W., Paczyński, B., & Venkat, R. 1994, Acta Astron., 44, 165

Udalski, A., et al. 1995, in preparation

Woźniak, P. R., & Stanek, K. Z., 1995, ApJ, submitted